\documentstyle[aps,twocolumn,pra,graphicx]{revtex}
\begin{document}                % INITIALIZE - DONT CHANGE %
\draft
\wideabs{
\author{A. I. Lvovsky}
\address{Fachbereich Physik, Universit\"at Konstanz}
\address{D-78457 Konstanz, Germany}
\author{J. H. Shapiro}
\address{Research Laboratory of Electronics}
\address{Massachusetts Institute of Technology}
\address{Cambridge, MA 02139}
\date{\today}
\title{Nonclassical character of statistical mixtures of the single-photon and
vacuum optical states}
\maketitle

\begin{abstract}
We demonstrate, theoretically and experimentally, that statistical
mixtures of the vacuum state $|0\rangle$ and the single-photon Fock
state $|1\rangle$ are nonclassical according to the Vogel criterion
(W. Vogel, Phys. Rev. Lett. {\bf 84}, 1849 (2000)), regardless of
the vacuum fraction. The ensembles are synthesized via
conditional measurements on biphotons generated by means of
parametric downconversion, and their quadrature statistics are
measured using balanced homodyne detection. A comparative review of
various quantum state nonclassicality criteria is presented.
\end{abstract}
 }

\section{Introduction} Nonclassical states of the electromagnetic
field have for many years provided an excellent playground for
testing fundamental concepts of quantum mechanics \cite{Review}.
Recently, nonclassical light has also become an important asset to
the rapidly developing applied fields of quantum optics, such as
quantum communication and quantum information technology. Its applications,
to name only a few, include quantum cryptography \cite{Gisin},
interferometric measurements \cite{DiMartini} and linear-optics
quantum computation \cite{KLM}.

Upon this background, it is becoming more important to have simple
criteria for identifying a particular quantum state as nonclassical.
This question has attracted strong interest from the first days of
quantum optics and a number of solutions, both for the single-mode
and multi-mode cases, have been proposed
\cite{MW,Dalton,Squeezing,osc,neg_Wig,nonclass_refs}. Although there
exists a commonly accepted formal definition of a nonclassical state,
no necessary and sufficient criterion has been proposed that would
allow verification of an optical state's nonclassical character in a
simple experiment.

For single-mode optical fields, nonclassical states are commonly defined as
those
that cannot be represented as a statistical mixture of coherent states.
This can be
reformulated in terms of the Glauber-Sudarshan $P$-function \cite{GS}: if
the latter
is positive definite, i.e., if it can be interpreted as a probability
density, then
the state possesses a classical analog.

The practical application of the above definition requires complete
information about the quantum state in question so that the $P$
function can be reconstructed. In experimental practice, however,
only partial information about a quantum state is normally available
and it is often degraded by detection noise and losses. The goal of a
nonclassicality criterion is to enable a conclusion about the
character of a quantum state based only on this partial information.
Known signatures of nonclassicality, frequently associated with a
particular method of quantum-state characterization, include
antibunching and sub-Poissonian photon statistics \cite{MW,Dalton},
squeezing \cite{Squeezing}, photon number oscillations \cite{osc},
and negative values of the Wigner function \cite{neg_Wig,Fock}.

All of these nonclassicality conditions are sufficient but not necessary
and leave wide classes of quantum states outside their scope. For
example, the Fock states are sub-Poissonian, but exhibit quadrature
noises above the shot noise level; phase-squeezed states are nonclassical,
but possess super-Poissonian photon statistics and positive Wigner
functions. This situation was recently improved by Vogel, who
showed that a quantum state is nonclassical if the absolute magnitude
of the state's characteristic function exceeds that of the vacuum
state at any point in inverse phase space \cite{Vogel}. This
modification of the traditional squeezing criterion covers a very
wide set of quantum optical states. As noticed by Di\'osi, however,
there are quantum states that are nonclassical but do not satisfy
the Vogel criterion, so the latter is still not a necessary condition
for nonclassicality \cite{Diosi}.

The Vogel criterion is now very relevant because of recent progress in
quantum homodyne tomography of highly nonclassical optical states.
In a recent experiment, Lvovsky and co-workers prepared the
single-photon Fock state $|1\rangle$ by conditional measurements on a
photon pair produced via parametric downconversion
\cite{Fock}.  The phase-averaged Wigner function reconstructed in the
measurement showed a strong dip around the phase-space origin,
reaching classically-impossible negative values.

Although negativity of the Wigner function is very strong indication
of a quantum state's nonclassical character, it covers many fewer
states than does the Vogel condition. As an example, we consider
statistical mixtures of the single-photon Fock state $|1\rangle$ and
the vacuum state $|0\rangle$:
\begin{equation}
\hat\rho_{\rm mix}=\eta|1\rangle\langle 1|+(1-\eta)|0\rangle\langle
0|,\label{rho_mix}
\end{equation}
for $0< \eta\le 1$. The Wigner function associated
with this mixed state takes on negative values only for $\eta>0.5$ \cite{Ou97}.
Moreover, reconstruction of the Wigner function requires
acquisition of a set of phase-stabilized marginal distributions
associated with different quadratures, or (as in \cite{Fock}) the
assumption that the Wigner function is rotationally invariant and hence
deducible
from a single phase-randomized marginal distribution. On the other hand, as we
demonstrate in this paper, the Vogel criterion establishes the nonclassical
character of ensembles (\ref{rho_mix}) for any $0<\eta\le 1$, and
it does so without requiring any assumptions regarding rotational symmetry or
local-oscillator phase stabilization.

\section{Homodyne detection and nonclassicality}
Vogel's nonclassicality condition is applicable to quantum-state data
obtained from balanced homodyne detection (BHD).  BHD is a technique for
performing
phase-sensitive measurements, i.e., field quadrature measurements, on a
light wave.
First proposed by Yuen and Chan in 1983
\cite{YuenChan}, it has become one of the main techniques for quantum-state
characterization. It has been used to detect squeezed states of the
electromagnetic field \cite{squeezing}, to quantify various
quantum optical states tomographically \cite{HomoTomo}, to
demonstrate Einstein-Podolsky-Rosen type quantum correlations
\cite{EPR}, and to carry out continuous-variable quantum teleportation
\cite{teleport}.

To perform BHD, the electromagnetic wave whose quantum state is to be
determined is overlapped on a beam splitter with a relatively strong
local oscillator (LO) wave in the same optical mode. The two fields
emerging from the beam splitter are incident on two high-efficiency
photodiodes whose output photocurrents are subtracted. The
photocurrent difference is proportional to the value of the electric
field quadrature operator $\hat E(\theta)$ in the signal mode, where
$\theta$ is the relative optical phase between the signal and LO.
Repeated homodyne measurements on the same quantum state yield the
statistical properties of the quantum noise associated with this
quadrature. A set of quadrature noise measurements acquired at
different values of $\theta$ is sufficient to reconstruct the Wigner
function and the density matrix of the quantum state
\cite{Leonhardt}.

To demonstrate the nonclassical character of quantum states measured
via BHD, comparison with the semiclassical theory of photodetection
can be used. According to this theory, light is a classical
electromagnetic wave and the detection noise arises from the discrete
nature of the charge carriers created in the detection process. Assuming
that BHD
is carried out in the time-resolved, pulsed regime, the number of
photoelectrons,
$N_i$, produced by detection of the $i$th optical pulse consists of two
independent
parts: the quadrature
measurement $N_{i,E}$ of the classical signal field plus the photoelectron shot
noise $\delta N_{i,S}$,
\begin{equation}
N_i= N_{i,E} + \delta N_{i,S}.
\label{deltaN}
\end{equation}

Assuming the signal field to be much weaker than the local
oscillator, we can regard the shot noise as being entirely due to the
LO.  A local oscillator pulse of energy $E_{\rm LO}$ generates an
average of $N_0/2=\eta E_{\rm LO}/2\hbar \omega$ charge carriers in
each photodiode, where $\eta$ is their quantum efficiency and
$\omega$ is the optical frequency. These charge carriers are subject
to Poisson statistics and carry a mean-square fluctuation of
$\langle\delta^2(N_0/2)\rangle=N_0/2$. When the two photocurrents are
subtracted, the sum of their independent shot noises leads to a total
mean-square fluctuation strength of $\langle\delta^2
N_{S}\rangle=N_0/2+N_0/2=N_0$ per pulse. According to the
semiclassical theory, therefore, the total mean-square photodetection
noise,
\begin{equation}
\langle\delta^2 N\rangle=\eta E_{\rm LO}/\hbar \omega+\langle\delta^2
N_E\rangle, \label{delta2N}
\end{equation}
always equals or exceeds the shot-noise level observed when the
signal field is vacuum ($N_E\equiv 0$).

Whenever the signal field is coherent, i.e., carries no classical
noise, the prediction (\ref{delta2N}) of the semiclassical
calculation is identical to that of the ``correct" theory, in which
both the detector and the light are treated quantum mechanically. In
keeping with standard practice in quantum optics, we define an
optical state to be {\it classical} if it is a statistical mixture of
coherent states.  Classical states, therefore, have BHD photocurrent
noise that can be correctly quantified using the above semiclassical
theory. A squeezed state, on the other hand, is an example of a
nonclassical state because its quadrature noise at some phase angles
falls below the shot-noise level.

A more general approach to nonclassicality has been proposed by Vogel,
whose treatment is re-derived in the following. Let ${\rm pr}(N)$ be the
probability
of detecting exactly $N$ photoelectrons in the BHD subtraction output. In
semiclassical photodetection theory, the optical quadrature noise
of the signal field is statistically independent of the charge-carrier shot
noise,
so there exist separate probability distributions
${\rm pr}_E(N)$ for the field noise and ${\rm pr_S}(N)$ for the shot noise.
According to Eq. (\ref{deltaN}), the measured
${\rm pr}(N)$ is then the convolution of these two distributions:
\begin{equation}
{\rm pr}(N)={\rm pr}_E(N)\otimes {\rm pr}_S(N). \label{conv}
\end{equation}
Because ${\rm pr}(N)$ and ${\rm pr}_S(N)$ can be determined
experimentally, the question of classicality reduces to the existence
of a probability distribution ${\rm pr}_E(N)$ that would satisfy
the above equation.

Taking the Fourier transform of Eq. (\ref{conv}), we find
\begin{equation}
{\rm F[pr]}(\nu)={\rm F[pr}_E](\nu)\,\times\, {\rm F[pr}_S](\nu),
\label{Fourier}
\end{equation}
where ${\rm F[pr]}$ denotes the Fourier transform of the marginal
distribution (i.e., the cross-section of the state's Wigner
characteristic function associated with the particular LO phase under
investigation) and $\nu$ is the transform variable \cite{footnote}.
From probability-distribution normalization, which implies that
$|{\rm F[pr}_E](\nu)|\le 1$ for all $\nu$, we obtain a new necessary
condition for the classicality of a quantum state --- the Vogel
criterion \cite{Vogel}:
\begin{equation}
|{\rm F[pr]}(\nu)|\le|{\rm F[pr}_S](\nu)|  \label{cond}
\end{equation}
for all $\nu$. If inequality (\ref {cond}) is invalid for {\it any}
quadrature at
{\it any} value of $\nu$, then equation (\ref{conv}) cannot be satisfied
and the
state is nonclassical.

An important feature of this nonclassicality condition (not mentioned
in the original Vogel paper) is that it can be generalized to
statistical mixtures of several marginal distributions associated
with different phases. In other words, there is no need to maintain a
stable phase relation between the LO and the optical state being
tested. As soon as the histogram of the quadrature measurements
acquired in an experimental run violates inequality (\ref {cond}),
the state is known to be nonclassical.

\section{Statistical mixtures of the vacuum and single-photon states}
\subsection{Theoretical analysis}
The ensembles defined by (\ref{rho_mix}) are completely characterized by the
Wigner function
\begin{equation}
W_\eta(X,P)={2 \over \pi} \left( {4\eta (X^2+P^2)+1-2\eta},
\right)e^{-2 (X^2+P^2)}
\end{equation}
which is equivalent to the following, phase-independent marginal
distribution for a
quadrature measurement:
\begin{equation}
{\rm pr}_\eta(X)=\sqrt{\frac{2}{\pi }}\,(1-\eta+4\eta
X^2)e^{-2\,X^2}.
\end{equation}

The mean-square quadrature deviation of the above distribution is
\begin{equation}
\langle\delta^2X\rangle=1/4+\eta/2\label{delta2X}\end{equation}
which always exceeds the vacuum-state value
of $1/4$. On the other hand, the
Fourier image of this distribution,
\begin{equation}
{\rm F[pr]_\eta}(\nu)= \int {\rm pr}_\eta(X)e^{i\nu X} dX=(1-\eta
\nu^2/4)e^{-\nu^2/8},
\end{equation}
has an absolute value that exceeds that of the vacuum state (${\rm
F[pr]_0}(\nu) = e^{-\nu^2/8}$) over two semi-infinite intervals of $\nu$
(Fig. 1). A
simple calculation shows that for a given $\eta$, the difference between the
characteristic functions $D_\eta(\nu)=|{\rm F[pr]_\eta}(\nu)|-|{\rm
F[pr]_0}(\nu)|$
reaches its maximum value of $D_\eta(\nu_{\rm opt})=2\eta
\exp[-(1+\eta)/\eta]$ at
\begin{equation}\nu_{\rm opt}^2=8(1+\eta)/\eta\label{nu_opt}.\end{equation}
For any $0<\eta\le 1$, the quantity $D_\eta(\nu_{\rm opt})$ is positive.
Ensembles (\ref{rho_mix}) are thus always nonclassical according to
the Vogel criterion, even though they are never squeezed and their Wigner
distribution is non-negative for $0\le \eta <1/2$.

\begin{figure}
\begin{center}
\includegraphics[width=0.45\textwidth]{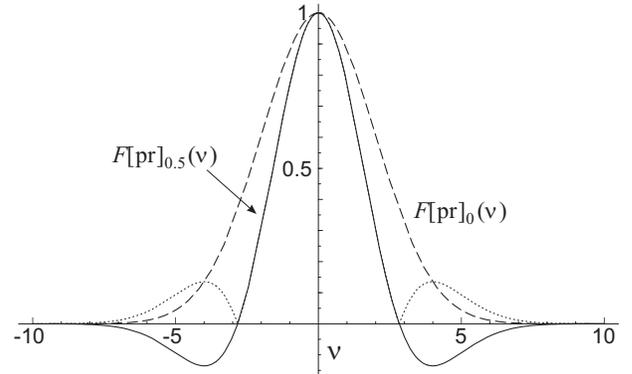}
\caption{\label{Fourier-th} Vogel nonclassicality of the state
$\hat\rho_{\rm mix}= \eta|1\rangle\langle 1|$
$+(1-\eta)|0\rangle\langle 0|$ with $\eta=0.5$. Cross-section of the
characteristic function of this state (solid line) is shown along
with that of the vacuum state (dashed line). The dotted line displays
the absolute value of ${\rm F[pr]_\eta}(\nu)$, which exceeds its
vacuum state counterpart over two semi-infinite intervals.}
\end{center}
\end{figure}

Experimentally, the characteristic function cross-section is
evaluated from a set of $N$ individual quadrature measurements $X_i$ as
an estimate (empirical average)
\begin{equation}{\tilde{\rm F}\rm [pr]_\eta}(\nu)=\langle e^{i\nu
X_i}\rangle_N.\label{avg}\end{equation} This evaluation suffers a
mean-square estimation error
\begin{equation} \langle |{\tilde{\rm F}[\rm pr]_\eta}(\nu)
- {\rm F[pr]_\eta}(\nu)|^2\rangle = \frac{1}{N}\left(1-
|{\rm F[pr]_\eta}(\nu)|^2\right).
\label{delta2F}\end{equation}

\begin{figure}
\begin{center}
\includegraphics[width=0.45\textwidth]{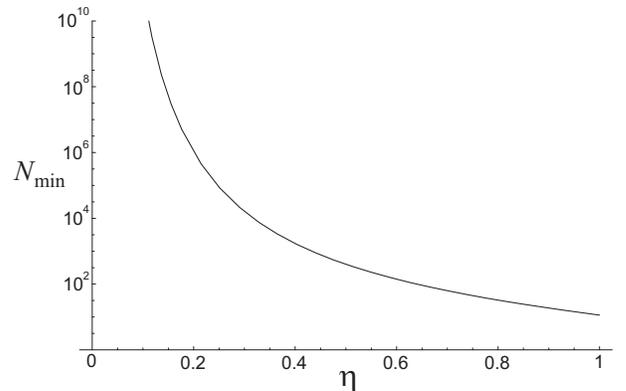}
\caption{\label{Nmin} Minimum number of quadrature samples that must
be acquired in order to establish the nonclassical character of the
state (\ref{rho_mix}) beyond the margin of statistical uncertainty. }
\end{center}
\end{figure}

The nonclassical character of a quantum state can be established only
if this uncertainty does not exceed $D_\eta(\nu_{\rm opt})$. Since
the latter function decays quickly with reducing $\eta$,
demonstrating the nonclassicality of ensembles (\ref{rho_mix}) at low
efficiencies requires an exponentially large number of samples to be
acquired (Fig. \ref{Nmin}).

\subsection{Experiment}

\begin{figure}
\begin{center}
\includegraphics[width=0.45\textwidth]{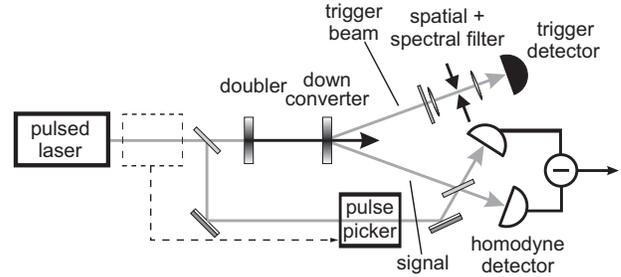}
\caption{\label{setup} Master scheme of the experimental setup. The
new position of the pulse picker allowed a much higher
photon-pair production rate to be achieved.}
\end{center}
\end{figure}

%%%JHS edit
%%%\paragraph{The setup}
%%%
The experimental apparatus employed in our work was almost identical
to that described in \cite{Fock} (Fig. \ref{setup}). The 790-nm,
1.6-ps output of a Spectra Physics Tsunami laser was frequency
doubled in a single pass through a 3-mm LBO crystal and then passed
on to a 3-mm BBO crystal for downconversion. The downconverter
operated in a type-I frequency degenerate, non-colinear
configuration. A single photon counter (EG\&G SPCM-AQ-131) was placed
in one of the emission channels --- labeled trigger --- to detect
photon-pair creation events and to trigger a homodyne system placed
in the other emission channel --- labeled signal. In this way, pulses
selected for homodyne measurements are only those for which a photon
has been emitted into the signal channel, thus preparing
single-photon Fock states by conditional measurements.

We used a small fraction of the original optical pulses from the
pulse picker --- split off before the frequency-doubler --- as the local
oscillator for the homodyne system. These pulses have to be
temporally and spatially mode-matched to the mode of the photons in
the signal channel \cite{MMpaper}. In order to resolve the quantum
noise of individual laser pulses, a time-domain homodyne detector
with ultra-low electronic noise ($\sim$$1000$ electrons per pulse),
high subtraction efficiency ($>$$83$ dB), and high frequency
bandwidth (DC -- 2 MHz) has been used \cite{FockHD}.

The apparatus employed in \cite{Fock} suffered a significant
drawback owing to its very low pair production rate (around 1 pair per 4
seconds). This was caused, in particular, by a deliberate factor-of-100
reduction
of the Ti:sapphire master laser's 82-MHz pulse repetition rate, which was
achieved by using an acousto-optical pulse picker. The necessity for this
reduction
arose from the relatively low ($\sim$1 MHz) bandwidth of the balanced homodyne
detector amplifier; the detector was unable to resolve the shot noise of
individual
laser pulses arriving at a higher rate.

In the present work, the pair generation rate was dramatically increased by
installing an electro-optical pulse picker in the beam path of the local
oscillator,
instead of right after the master laser. The entire setup, apart from the
homodyne
detector, was thus operating at the original laser repetition rate of 82 MHz.
Whenever a trigger photon detection event occurred, the optical shutter in
the local
oscillator opened to transmit a single local oscillator pulse that activated a
balanced homodyne measurement. In order to compensate for the trigger delay of
the optical shutter, a 50-ns optical delay line was introduced. The
replacement of the pulse picker and a number of other modifications
allowed us to enhance the pair production rate by a factor of more
than a 1000 in comparison to the work \cite{Fock}; the original
experiment that lasted 14 hours would only last 45 seconds with the
new apparatus.

Various experimental imperfections cause an admixture of the vacuum
$|0\rangle$ to the Fock state $|1\rangle$ that would be measured if
the setup were ideal \cite{Fock}. When optimally aligned, our
apparatus constructed ensembles $\hat{\rho}_{\rm mix}$ with the
highest overall efficiency $\eta = 0.61$. Further reductions of
$\eta$ --- to probe the Vogel criterion for $\hat{\rho}_{\rm mix}$ at
high vacuum fraction --- have been achieved by misaligning the
temporal synchronization between the LO and single-photon pulses.

\begin{figure}
\begin{center}
\includegraphics[width=0.42\textwidth]{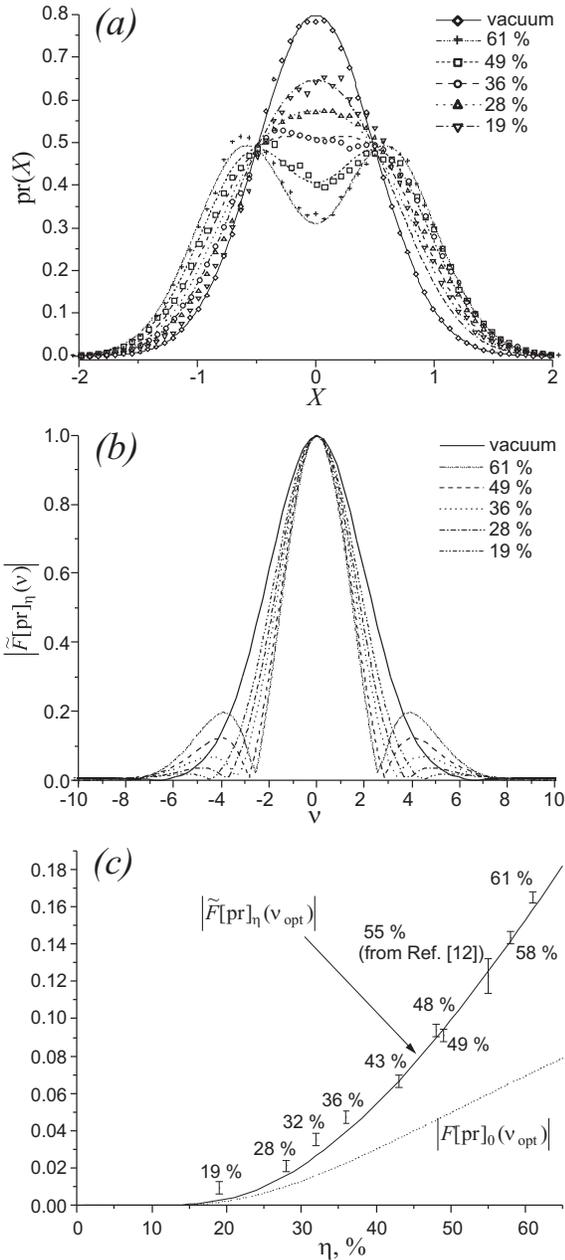}
\caption{\label{results} Experimental results. (a) Marginal
distributions associated with the vacuum state and ensembles
(\ref{rho_mix}) for four different efficiency values; (b)
Cross-sections of the characteristic functions obtained through
Fourier analysis of the homodyne data; (c) Theoretical (solid line)
and experimental values of $|{\rm\tilde F[pr]}_\eta(\nu_{\rm opt})|$.
The function $|{\rm F[pr]}_0(\nu_{\rm opt})|$ (dashed line), which
establishes the classical limit, is also displayed.}
\end{center}
\end{figure}

We acquired 9 data sets, each containing 100,000 electric field
quadrature measurements for the states (\ref{rho_mix}) with the
measurement efficiencies $\eta$ between 0.19 and 0.61 and a single
data set of 200,000 points for the vacuum state. The phase-averaged
Wigner functions reconstructed from the data sets corresponding to
$\eta=0.58$ and $\eta=0.61$, as expected, exhibited negativities
around the phase space origin point. All data sets were then binned
up to obtain their histograms (Fig. \ref{results} (a)). The Fourier
images $|{\rm \tilde F[pr]_\eta}(\nu)|$ of the marginal
distributions, calculated using Eq.\,(\ref{avg}) and shown in Fig.
\ref{results}(b), clearly violate inequality (\ref{cond}). The
nonclassicality of the states investigated is made manifest in Fig.
\ref{results}(c), where the values of $|{\rm \tilde
F[pr]_\eta}(\nu_{\rm opt})|$, reflecting the highest difference
between the classical and quantum behavior, are plotted.

Because the setup was not interferometrically stable, each of the
quadrature data sets corresponds to the superposition of marginal
distributions of ensembles (\ref{rho_mix}) associated with different
phases. This circumstance does not compromise our conclusion on the
nonclassical character of these states, as the latter does not rely on any
assumptions regarding phase stability or randomness.

Of special interest are the quadrature data acquired for the 0.28 and
0.19 efficiency values. The marginal distributions associated with
these efficiencies (Fig. \ref{results}(a)) are not only wider than
that of the vacuum state but also lack any fine structure which would
allow a direct decision on their nonclassical character.  Only the
Fourier analysis of these distributions allows us to draw such a
conclusion. These examples show that the suggested reformulation of
Vogel's criterion, stating that ``a quantum state has no classical
counterpart when these [marginal distribution] functions show
structures that are narrower than the corresponding distributions of
the ground state of the oscillator" \cite{Vogel} is not always
rigorous.

\section{Discussion}
Comparing the Vogel criterion and the negativity of the Wigner
function as nonclassicality conditions, we first notice that
fulfillment of either one does not automatically imply satisfaction
of the other. In the previous section we have considered quantum
states that are nonclassical according to the Vogel criterion but
have positive-definite Wigner functions. As an opposite example one
can use the ensemble presented in the Di\'{o}si correspondence on the
Vogel paper \cite{Diosi}:
\begin{equation}
\hat\rho = \sum_{n=1}^{\infty} 2^{-n}|n\rangle\langle n|.
\end{equation}
The Glauber-Sudarshan $P$-function for this state is
\begin{equation}
P_{\rm GS}(X,P) = \frac{2}{\pi}e^{-(X^2+P^2)}- \delta(X)\delta(P).
\end{equation}
Convolving this $P$-distribution with a circularly-symmetric
zero-mean, variance $1/4$ 2-D Gaussian function gives the Wigner
distribution \cite{Leonhardt},
\begin{equation}
W(X,P) = \frac{4}{3\pi}e^{-2(X^2+P^2)/3}
-\frac{2}{\pi}e^{-2(X^2+P^2)}.
\end{equation}
This distribution is negative at $X=P=0$, but (as shown by Di\'{o}si) the
above state fails the Vogel criterion for nonclassicality.

It is also instructive to compare the two nonclassicality criteria
from a purely classical viewpoint. To what extent can the fulfillment
of each condition persuade someone who only believes in classical
physics that something is wrong with her picture of the world? In
this aspect, the two criteria are strikingly different in terms of
fundamental assumptions the ``classicist" must maintain in order to
be convinced of a contradiction by observing one or the other
criterion fulfilled. The Vogel condition is based on the
semiclassical theory of photodetection, i.e., it assumes that
homodyne detection results in field-quadrature observation that is
embedded in a certain, well-defined amount of (shot) noise. Homodyne
tomography, from which the Wigner distribution in reconstructed,
makes no such postulate, and can be fully interpreted in the
framework of classical physics. In particular, this classical
interpretation makes the Wigner distribution a (positive definite)
classical probability distribution over phase space. All that is
presumed in making this classical interpretation are Maxwell's
equations for the electromagnetic wave and the fact that a
photodiode's output current is proportional to the intensity of the
incident beam (perhaps with some added classical noise). The
combination of a classically understandable measurement method and an
evident non-classicality of its result --- the negative Wigner
function, such as in the work \cite{Fock} --- provides very strong
evidence of quantum mechanics.

In the framework of this discussion it is interesting to compare the
Wigner function's negativity with another well-known signature of
nonclassicality, namely, the violation of Bell's inequality. In the
latter, the assumptions that are made are even less restrictive.
Causality is the only physical postulate; no assumptions need to be
made regarding the physical nature of the experiment itself. If a
tomographic reconstruction of a Wigner function with negative values
would convince a {\it physicist} that the world is quantum, a
loophole-free violation of the Bell inequality would convince a {\it
philosopher} who has no knowledge of physics whatsoever. From a
purely philosophical point of view, tomographic reconstruction of a
negative Wigner function is stronger evidence of quantum mechanics
than is the incompatibility of the detection noise distribution with
the semiclassical photodetection theory, but weaker than violation of
Bell's inequalities.

In terms of experimental simplicity, however, the sequence goes the
other way. Apart from the fact that quantum states with negative
Wigner functions (e.g., Fock states) are quite exotic and difficult
to synthesize, the complete reconstruction of the Wigner distribution
requires acquisition of a full set of marginal distributions with
fixed relative phase between the LO and the state being examined. On
the other hand, verification of the Vogel criterion can be applied to
a single marginal distribution or to a superposition of marginals
associated with different phases. To date, no loophole-free violation
of Bell's inequalities has been reported \cite{Bell}.

\section{conclusions}
We have shown that the electric field quadrature statistics of the
ensembles (\ref{rho_mix}) of the vacuum state $|0\rangle$ and the
single-photon Fock state $|1\rangle$
%%%JHS edit
%%%is
%%%
are
nonclassical according to
the Vogel criterion \cite{Vogel}. These states were synthesized and
measured in a setting similar to that of Lvovsky {\it et al.}
\cite{Fock} but with an improved pair-production rate achieved thanks
to a new method of pulse picking. The Vogel criterion has been
generalized to apply to quadrature distributions obtained in
interferometrically unstable settings. A quantitative analysis of
statistical errors has been given.

The authors are grateful to Prof. W. Vogel for helpful discussions.
A. L. is sponsored by the Alexander von Humboldt foundation. He also
thanks Prof. J. Mlynek for support. The research of J. H. S. was
supported by the U. S. Army Research Office. Please address all
correspondence to Alex.Lvovsky@uni-konstanz.de.


\begin{references}
\bibitem{footnote} The function ${\rm F[pr]}(\nu)$ used in our paper corresponds to
the function $G(k,\phi)$ used by Vogel \cite{Vogel}. Our
Eq.\,(\ref{Fourier}) is essentially identical to  Eq.\,(9) in
\cite{Vogel}, except for the quadrature scaling convention which is a
factor of 2 different.
\end{references}
\end{document}